\documentclass[preprint,3p]{elsarticle}
\usepackage[table]{xcolor}
\usepackage[english]{babel}
\usepackage[utf8]{inputenc}
\usepackage{eurosym}
\usepackage{float}
\usepackage{flushend}
\usepackage{setspace}
\usepackage{amssymb}
\usepackage{tikz}
\usetikzlibrary{fit}
\usetikzlibrary{backgrounds}
\usetikzlibrary{positioning,fit,calc}
\usepackage{bbding}
\usepackage[]{tabularx}
\usepackage{circuitikz}

\usepackage{booktabs,multirow}

\usepackage{moreverb}
\usepackage[colorlinks,bookmarksopen,bookmarksnumbered,citecolor=red,urlcolor=red]{hyperref}
\usepackage{enumerate}
\usepackage{textcomp}
\usepackage{subcaption}
\usepackage{graphicx}
\usepackage{amssymb}
\usepackage{amsthm}
\usepackage{amsmath}
\usepackage{multirow}
\usepackage{array}
\usepackage{bm}
\usetikzlibrary{arrows}
\usepackage{enumitem}
\usepackage{arydshln}
\usepackage{color, colortbl}

\begin{document}
	\begin{frontmatter}

		\title{A new approach to single-phase  systems under sinusoidal and non-sinusoidal supply using   geometric algebra}

		\author[esi]{Francisco G. Montoya}
		\ead{pagilm@ual.es}
		\author[esi]{Raúl Baños}
		\ead{rbanos@ual.es}
		\author[esi]{Alfredo Alcayde}
		\ead{aalcayde@ual.es}
		\author[esi]{Francisco M. Arrabal-Campos}
		\ead{fmarrabal@ual.es}
		
		\address[esi]{CeiA3, Department of Engineering, University of Almería, Carretera de Sacramento s/n, 04120 Almería (Spain)}

\begin{abstract} 

The aim of this work is to present major upgrades to existing power theories based on geometric algebra for single-phase circuits in the frequency domain. It also embodies an interesting new approach with respect to traditionally accepted power theories, revisiting power concepts in both sinusoidal and non-sinusoidal systems with linear and nonlinear loads for a proper identification of its components to achieve passive compensation of true non-active current. Moreover, it outlines traditional power theories based on the apparent power $S$ and confirms that these should definitively be reconsidered. It is evidenced that traditional proposals based on the concepts of Budeanu, Fryze and others fail to identify the interactions between voltage and current harmonics. Based on the initial work of Castro-Núñez and others, new aspects not previously included are detailed, modified and reformulated. As a result, it is now possible to analyze non sinusoidal electrical circuits, establishing power balances that comply with the principle of energy conservation, and achieving optimal compensation scenarios with both passive and active elements in linear and non-linear loads.

\end{abstract}    


\end{frontmatter}

\section{Introduction}

The study of power flow in electrical systems is a century-old issue. Engineers and scientists around the world have debated about it throughout the XX century and up to now. In sinusoidal and linear systems, there is a clear consensus that electrical power can be analysed through a decomposition that takes into account the average power over a period of time $T$, namely the \textit{active power} $P$, in addition to a quadrature term, the \textit{reactive power} $Q$. However, in non-sinusoidal systems, there are disagreements among researchers. Several schools have emerged around this topic, each having a different interpretation depending on the approach. In Table \ref{tab:power_contributions}, relevant theories and their contributions are summarised \cite{castro2013use}.

\begin{table*}[]
	\centering
	\begin{tabular}{@{}ll@{}}
		\toprule
		\multicolumn{1}{c}{\textbf{Author}}                    & \multicolumn{1}{c}{\textbf{Contribution}}                                   \\ \midrule
		\textbf{Nabae and Tanaka}                        & Powers based on instantaneous space vector                              \\
		\textbf{Shepherd and Zakikhani}                  & Definition of reactive power                                            \\
		\textbf{Kusters and Moore}                & Inductive and capacitive current                                        \\
		\textbf{Depenbrock}                                 & Fryze-Buchholz-Depenbrock (FBD) Power Theory                            \\
		\textbf{Sharon}                                     & Reactive power definitions                                              \\
		\textbf{Slonim and Van Wyck}                 & Definition of active, reactive and apparent powers                      \\
		\textbf{Emanuel}                                 & Definitions of apparent power                                           \\
		\textbf{Czarnecki}								& Current's Physical components \\
		\textbf{Peng and Lai}                        & Generalized instantaneous reactive power theory                         \\
		\textbf{Ferrero and Superti-Furga}                        & The Park power theory                                                   \\
		\textbf{Rossetto and Tenti}                            & Instantaneous orthogonal currents                                       \\
		\textbf{Peng}                                          & Generalized non-active power theory                                     \\
		\textbf{Willems}                                       & Instantaneous voltage and current vectors                               \\
		\textbf{Fillipski}                               & Elucidation of apparent power and power factor                          \\
		\textbf{Watanabe}                                & Generalised theory of instantaneous powers $\alpha$-$\beta$-0 transformation         \\
		\textbf{LaWhite and Ilic}                       & Vector space decomposition of reactive power                            \\
		\textbf{Ghassemi}                                   & Definition of apparent power based on modified voltage                  \\
		\textbf{Cohen and deLeon}         & Time-domain representation of powers                                    \\
		\textbf{Zhang}                                         & Universal instantaneous power theory                                    \\
		\textbf{Lev-Ari and Stankovic}                & Reactive power definition via local Fourier transform                   \\
		\textbf{Haque}                                   & Single phase PQ theory                                                \\
		\textbf{Menti and Zacharias} & Introduced Geometric Algebra to non-sinusoidal power theory             \\
		\textbf{Castilla and Bravo}               & Extended the use of Geometric Algebra in non-sinusoidal power theory     \\
		\textbf{Xianzhong and Guohai}             & Generalised theory of instantaneous reactive power for multiphase system \\
		\textbf{Shin-Kuan  and Chang}                  & Instantaneous power theory based on active filter                        \\
		\textbf{Dalgerti}                                & Concepts based on instantaneous complex power approach                  \\ \bottomrule
	\end{tabular}
\caption{Contributions to power theory by main authors. Reproduced from \cite{castro2013use}.}
\label{tab:power_contributions}
\end{table*}

There are two primary approaches for power theory, i.e., time-domain and frequency-domain. The former has had a significant impact, especially in three-phase systems, and has a very specific goal, i.e., compensation through active filters. The second one (in its various forms) has been widely used in different  electrical systems, not only for reactive power compensation, but also for electric circuit analysis, power quality analysis, etc. %This is why ideas in the frequency domain attract greater interest because they affect much more systems than time-domain theories. 
In addition, some time-domain theories (Akagi-Watanabe \cite{akagi2017instantaneous}) have been criticised  \cite{czarnecki2004some} because they cannot satisfactory explain the energy exchange processes in non-sinusoidal and distorted situations. Based on the above motivations, frequency-domain theories have gained a great attraction by the scientific community. However, there are two important common factors  to all these theories that should be examined in more detail:

\begin{enumerate}
	\item The majority uses the apparent power concept as a result  of RMS voltage $V$ and  current $I$ product, i.e., $S=VI$.
	\item Most of them are supported by complex number algebra, $\bm{S}=\bm{VI^{*}}$, where $\bm{V}$ is the voltage phasor, $\bm{I^*}$ is the conjugated current phasor and $\bm{S}$ is the complex apparent power.    
\end{enumerate}  

The complex apparent power arises from a well-known definition that has been traditionally accepted by the community from its inception. It intends (through a pretended analogy with the instantaneous power $p(t)=v(t)i(t)$) to universalise a term that numerous studies have shown does not represent any physical quantity. Moreover, it does not meet the basic principle of conservation of energy (PCoE) \cite{czarnecki1987wrong,czarnecki1985considerations,czarnecki2004some,filipski1993apparent}. That it delivers correct mathematical results in sinusoidal systems does not imply that it can be properly generalised to non-sinusoidal systems. However, it remains so popular because of the usual decomposition of the current in quadrature terms. It is suggested that the  formulation based on currents could be a better representation from an electrical engineering point of view. For example, the current physical components (CPC) theory \cite{czarnecki2008currents} states that the current in single-phase systems can be decomposed as follows:

\begin{equation}
	i(t)=i_a(t)+i_s(t)+i_r(t)+i_G(t)
\end{equation}

where $i_a(t)$ is the active or Fryze current, $i_s(t)$ is the scattered current, $i_r(t)$ is the reactive current and $i_G(t)$ is the load harmonic generated current. As demonstrated by Czarnecki, all of the above terms are in cuadrature, so the RMS values fulfill: 

\begin{equation}
	\|I\|^2=\|I_a\|^2+\|I_s\|^2+\|I_r\|^2+\|I_G\|^2
	\label{eq:curr_quad}
\end{equation}

Note that the Euclidean norm is used. If equation (\ref{eq:curr_quad}) is multiplied by the squared voltage, it results in the following:

\begin{equation}
\begin{aligned}
S^2=\|V\|^2\|I\|^2&=\|V\|^2\|I_a\|^2+\|V\|^2\|I_s\|^2+\|V\|^2\|I_r\|^2+\|V\|^2\|I_G\|^2=P^2+D^2_s+Q^2_r+D^2_G
\label{eq:pot_cuad}
\end{aligned}
\end{equation}

Equation (\ref{eq:pot_cuad}) suggests that a power balance is achieved, and the apparent power could be composed of other powers caused by certain physical processes. However, this is questionable because the derivation of (\ref{eq:pot_cuad}) from (\ref{eq:curr_quad}) cannot disguise the physical reality of the problem:  current $I$ can be decomposed into certain components following Kirchhoff’s current laws and  those components are orthogonal to each other. However, the above does not necessarily entail that the derived power terms  generate a valid decomposition  because:
\begin{equation}
	S\ne P+D_s+Q_r+D_g
\end{equation} 

The true physical meaning of the electrical power is in $p(t)$, i.e., the instantaneous power expressed as energy transferred per  unit time as follows:

\begin{equation}
	p(t)=\frac{dW}{dt}
\end{equation}

and its mean  value $P$, which represents the average power demanded by a load during a time period of $T$ and transformed into useful work, as follows: 

\begin{equation}
	P=\frac{1}{T}\int_{0}^{T}p(t)dt
\end{equation}

The efforts of scientists and engineers for years have focused on explaining the difference between $P$ and $S$ through various decompositions based on the quadrature of artificial power components, somewhat unsuccessfully regarding the physical meaning, beyond some mathematical parallelism.  In our humble opinion, a proper decomposition of the current should be favoured instead. Even though the multiplication of RMS current and voltage results in a mathematical concept of multi-component power, at the energy level, the active power $P$ is the ultimate objective. The rest is somehow less irrelevant because it does not contribute to any useful work. Furthermore, from an economic point of view, the decomposition of the current is where we should focus our efforts to detect that part that can be eliminated or compensated  locally to avoid energy losses in power lines or a degradation of power quality. It is therefore essential to find appropriate tools or  procedures that fully and accurately describe its calculation  and also being consistent with the laws that govern  electrical circuits.

This distinction between mathematics and physics has already been noted repeatedly by several authors \cite{orts2011discussion,czarnecki2004some,czarnecki1987wrong,willems2011budeanu}. Mathematical correction certainly is an essential requirement for a theory to be accepted by the community. However, there are some authors that have considered its physical interpretation as a necessary condition. It is our contention to highlight the attempt to assign physical meaning to the non-active power, with very little success at the moment. Only the active power $P$ and instantaneous power $p(t)$ have been proved to manifest full physical sense. Therefore, it is worth to develope a decomposition of the current into components that may be useful for specific objectives, such as the compensation of the non-active current or the improvement of power quality.           

In the quest for tools,  methods or theories that adequately describe power exchange processes between source and load in any type of system, the geometric algebra (GA) presented by Clifford in 1878, supported by the work of Hamilton (1843) and Grassmann (1844), and then recovered by Hestenes in the 1960's \cite{hestenes2012clifford,hestenes2012new}, is the one that fills the gaps detected in the algebra of complex numbers. Several studies have demonstrated the success of GA in disciplines such as relativistic physics, electromagnetism and computer vision \cite{dorst2010geometric,chappell2014geometric,bernard1989multivectors}. In addition, the development of the theory of electrical power based on GA provides a new approach for solving power flow in electrical systems  because of its flexibility and capability to represent the multi-component nature of power  in sinusoidal and non-sinusoidal circuits. Specifically, the studies by Menti \cite{menti2007geometric}, Castro-Nuñez (C-N) \cite{castro2012ieee,castro2012advantages,castro2016m,castro2019theorems}, Montoya \cite{montoya2019quadrature,montoya2018power}, Castilla and Bravo \cite{castilla2008clifford,castilla2008geometric,castillahijo}, Lev-Ari \cite{lev2009geometric} and Petroianu \cite{petroianu2015geometric} demonstrate the capabilities of GA in the analysis of power systems. Based on this approach, a better understanding of power balances can be obtained and, more importantly, compliance with the conservation of energy principle is guaranteed, i.e., Tellegen's theorem is satisfied.

As shown in previous studies \cite{castro2010use,castro2016m,montoya2018power}, GA applied to sinusoidal, non-sinusoidal, linear and nonlinear circuits is a suitable technique to describe power flow in terms of the energy conservation principle. Thus far, the definition of geometric apparent power $\bm{M}$ as the geometric product of current and voltage as follows:

\begin{equation}
	\bm{M}=\bm{VI}=P+\bm{CN}=P+\bm{CN}_d+\bm{CN}_r
\end{equation}

has been proposed as an efficient method to describe the net power flow in electrical circuits. In the above equation, $\bm{CN}$ (geometric non-active power), $\bm{CN}_d$ (geometric distorted power) and $\bm{CN}_r$ (geometric reactive power) represent Clifford numbers. Additionally, the decomposition of the current into in-phase and quadrature components as follows:

\begin{equation}
	\bm{I}=\bm{I}_g+\bm{I}_b
\end{equation}   

has contributed to the development of methods for quadrature RMS current compensation \cite{montoya2019quadrature}. However, some shortcomings in the existing power formulation have been detected and this work is a contribution to propose new several ideas that help to improve and promote the right use of GA in power systems.

The remainder of this paper is structured as follows. In Section 2, the main contributions of this work are highlighted. In Section 3, the theoretical basis of most well-known power theories is summarised; in Section 4, the basic concepts of GA are introduced; in Section 5, the proposed analysis of electrical circuits by GA is reviewed; in Section 6, the new formulation is presented; and in Section 7, the main conclusions of this work are presented.

\section{Main contributions}

The flaws detected in a major GA power theory based on Castro-Núñez  is analysed and several corrections are proposed. 
\begin{itemize}
	\item The core of C-N theory is based on the geometric apparent power formulation $\bm{M}=\bm{VI}$. This work proofs that the right expression should be $\bm{M}=\bm{V}\bm{I}^{\dagger}$. This new formulation avoids two main drawbacks in C-N theory: i) the introduction of an artificial  correction factor $f(-1)^{\frac{k(k-1)}{2}}$ and ii) the proper calculation of non-active power terms, which otherwise are  miscalculated.
	\item The definition of sub- and inter-harmonics for power computations. In this way, a more comprehensive approach is used for power calculations.
	\item The decomposition of currents into an in-phase component $\bm{I}_g$ and a quadrature component $\bm{I}_b$ has been extended since it does not provide information for the minimum current necessary to deliver the same power $P$ of the original circuit. The so-called geometric Fryze current is then introduced. 
\end{itemize}

\section{Brief review of the main Power Theories}
The power theories that have historically influenced electrical engineers the most are briefly reviewed. The purpose is to put into context the contributions of each theory and, more importantly, show some flaws and thus highlight why it should carefully used.            

\begin{itemize}
\item \textbf{Budeanu's theory}. This is the theory that may have had the greatest impact historically and which any electrical engineer knows and learned in college. It was formulated in 1927 in the frequency domain and establishes the following:         
	
	\begin{equation}
	\begin{aligned}
		S^2&=\|V\|^2\|I\|^2=P^2+Q^2+D^2 \\
		P&=\sum_{n}V_nI_n\cos\varphi\\
		Q&=\sum_{n}V_nI_n\sin\varphi\\
		D&=\sqrt{S^2-P^2-Q^2}
	\end{aligned}
	\end{equation}
	
		The primary issue is that the reactive power $Q$ is not always clearly associated with a physical phenomena in distorted and non-linear circuits, as suggested in \cite{czarnecki1987wrong,czarnecki1997budeanu}. Similarly, the power $D$ does not represent a physical phenomenon; it is implicitly defined as a function of $S$. It is also not possible to correctly compensate the system to minimise the amount of current consumed to maintain a constant $P$.
	
		\item \textbf{Fryze’s power theory}. Formulated in 1931 in the time-domain, this theory introduces the important concept of active current, i.e., the minimum current needed to produce the power $P$ demanded by the load.
	\begin{equation}
	\begin{aligned}
		i(t)&=i_a(t)+i_F(t) \\
		S^2&=P^2+Q_F^2
	\end{aligned}
	\end{equation}

	Even though it contributes to the decomposition of active current $i_a(t)$ and in quadrature $i_F(t)$, the non-active power $Q_F$ lacks physical meaning and cannot adequately compensate  non-sinusoidal systems.

\item \textbf{Shepherd and Zakikhani's theory}. Formulated in 1972 in the frequency domain, this theory was the first to provide a reactive current $i_r(t)$ definition according to the concepts of the voltage and the current in quadrature. The theory provides an effective reduction of the current in the presence of harmonics through an optimal compensator. Unfortunately, this theory does not include the concept of active power $P$, and therefore, it lacks usability.    

\item \textbf{Akagi-Nabae Instantaneous Reactive Power (IRP) theory}. Introduced in 1984 and defined in the time domain, this theory asserts the possibility of effective current compensation using active elements based on power electronics. To do this, the theory applies the Clarke transform to switch from an $a$-$b$-$c$ three-phase system to an $\alpha$-$\beta$-0 stationary system. This theory has been analyzed in some studies like \cite{czarnecki2004some}, where  critical inconsistencies in the compensation of nonsinusoidal circuits are showed. Its scope, therefore, is limited to sinusoidal circuits under specific conditions.            

\item \textbf{Depenbrock's FBD (Fryze-Buchholz-Depenbrock) theory}. A time-domain theory formulated by Depenbrock in 1993, it generalises the concepts of Fryze into three-phase systems, but it is still based on the non-conservative concept of apparent power $S$, and therefore fails to satisfy Tellegen's theorem.

\item \textbf{Czarnecki's CPC theory}. This theory was formulated by Czarnecki in 1984 in the frequency domain. According to this theory, the current is decomposed into various quadrature components as follows:  

\begin{equation}
	i(t)=i_a(t)+i_r(t)+i_s(t)+i_G(t)
\end{equation}

where $i_a(t)$ is Fryze’s current, $i_r(t)$ is Shepherd’s current, $i_s(t)$ is the scattered current caused by the fluctuation of the conductance of the load with frequency and $i_G(t)$ is the harmonic current generated by the load. Therefore, the apparent power is as follows:

\begin{equation}
	S^2=P^2+D_s^2+Q^2+D_g^2
\end{equation}

The main contribution of the CPC theory is related to the identification of a fictitious current $i_s(t)$ and the supposed  physical meaning it gives to the decomposition of the currents. Unfortunately, this theory relies on the concept of apparent power $S$; consequently, it cannot avoid the disadvantages inherent in that proposition. The magnitude of reactive power does not have a specific meaning; therefore, it is not possible to verify the balance of powers. Additionally, the CPC theory does not allow a complete balance of currents and powers in each branch of the circuit. Recent works \cite{jeltsema2018physical} has also challenged this theory using linear time varying circuits.     
\end{itemize}

\section{Basics of Geometric Algebra}

The beginnings of GA date back to the XIX century with the studies conducted by Grassmann, Hamilton and especially Clifford. Despite the little significance it had at the time (attributed to the untimely death of Clifford), at present, it is regarded as a \textit{unified language for physics and mathematics} \cite{hestenes2012new}. The essence of GA lies in the notion of an invertible (geometric) product that captures the geometric relation between two vectors, i.e., the relation between their modules and the angle they form \cite{sangston2016geometry}. Many studies \cite{hestenes2012clifford,chappell2014geometric} have demonstrated that GA, when applied to physics and engineering problems, provides analysis tools far superior to those derived from vector calculus as proposed by Gibbs and traditionally applied. For example, complex number algebra, quaternions or even vectors have been proven to be members of GA subspaces. A very interesting ability of GA is that the properties and operators are easily applicable to spaces with any number of dimensions.

The basic principles of GA derive from widely established vector concepts. For example, a vector $\bm{a}=\alpha_1\bm{\sigma}_1+\alpha_2\bm{\sigma}_2$ (that has orientation, sense and magnitude) can be multiplied by another vector $\bm{b}=\beta_1\bm{\sigma}_1+\beta_2\bm{\sigma}_2$ in various ways, such that the result has different meanings. Equation (\ref{eq:inner}) defines the inner or scalar product, and the result is a scalar. Note again, that an Euclidean norm is used.

\begin{figure}[]
	\centering
	\includegraphics[width=0.5\textwidth]{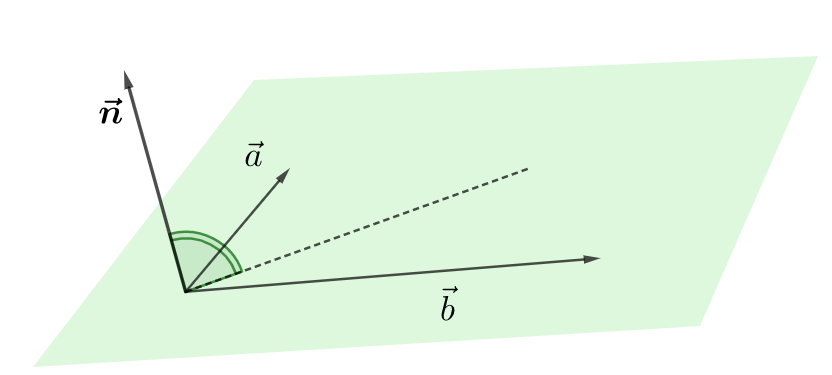}
	\caption{Classical vector product of vectors $\bm{b}$ and $\bm{a}$. The result is a vector $\bm{n}$, perpendicular to the plane formed by $\bm{b}$ and $\bm{a}$.}
	\label{fig:prodvectorial}
\end{figure} 

\begin{equation}
\bm{a}\cdot\bm{b}=\|\bm{a}\|\|\bm{b}\|\cos{\varphi}=\sum{\alpha_i\beta_i}
\label{eq:inner}
\end{equation}

Equation (\ref{eq:outer}) defines Grassmann’s product or the wedge product.

\begin{equation}
\bm{a}\wedge\bm{b}=\|\bm{a}\|\|\bm{b}\|\sin{\varphi}\ \bm{\sigma}_1\bm{\sigma}_2
\label{eq:outer}
\end{equation}

This product differs from the traditional vector or outer product (as in Figure \ref{fig:prodvectorial})  primarily because the result is neither a scalar nor a vector, but a new object termed \textbf{bivector}. The bivector is a key concept in GA and does not exist in linear algebra or traditional vector calculus \footnote{J.W. Gibbs also used the term \textit{bivector} but with a completely different meaning. See pag. 428 of the book "Vector Analysis" by E.B. Wilson in 1901}. GA demonstrates that the Gibbs vector product in 3D is simply the dual of the bivector \cite{chappell2014geometric}.

\begin{figure}[H]
	\centering
	\includegraphics[width=0.5\textwidth]{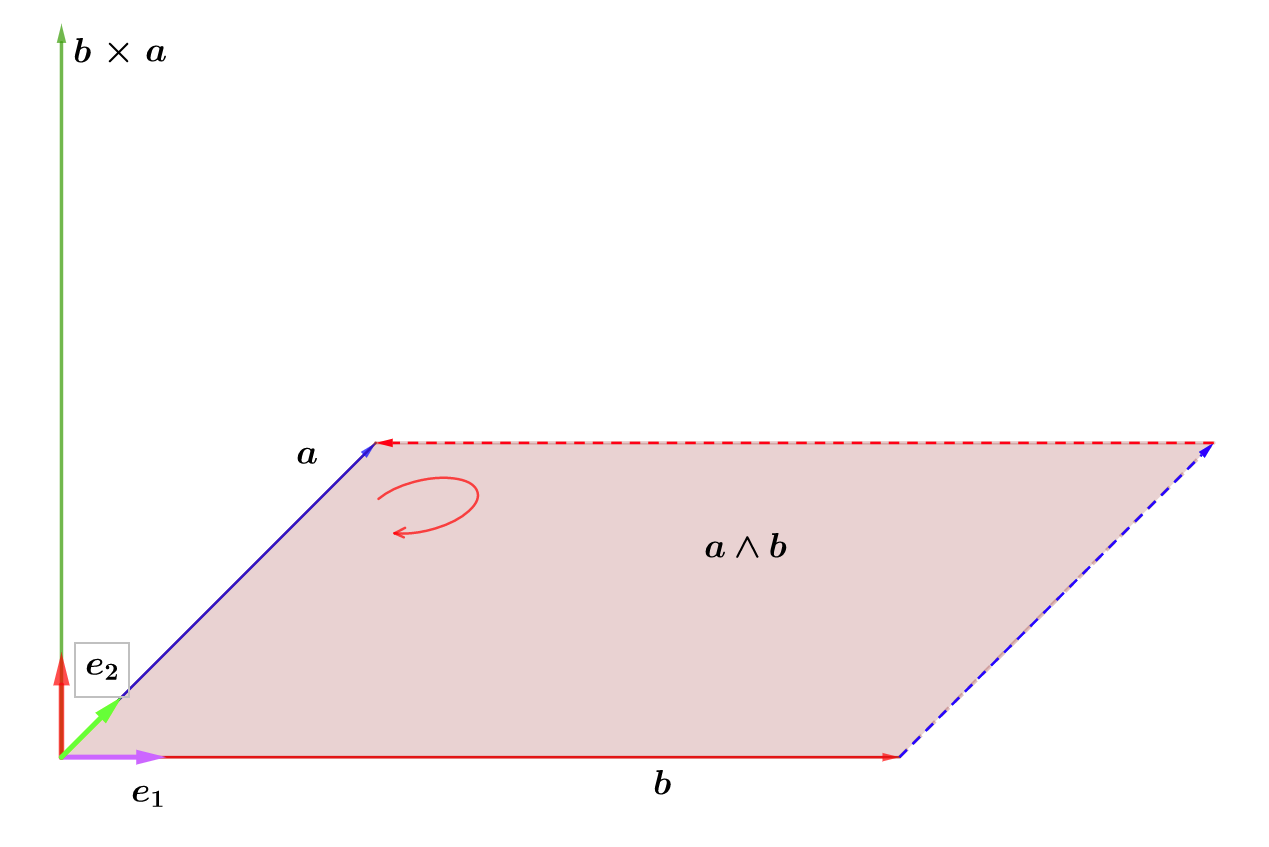}
	\caption{Representation of a bivector $\bm{a}\wedge \bm{b}$.}
	\label{fig:bivec}
\end{figure}

Similar to vectors, a bivector has orientation, sense and magnitude. Specifically, the area defined by vectors $\bm{a}$ and $\bm{b}$ is the geometric representation of the bivector (see Figure \ref{fig:bivec}) while the oriented arc represents the sense. An essential property of the wedge product is that it is anticommutative, i.e., $ \bm{a} \wedge \bm{b} = -\bm{b} \wedge \bm{a}$. Based on the above definitions, bivectors operate similar to vectors, i.e., bivectors can be added, multiplied and their inverse can be found. Similar to vectors, bivectors can also be expressed as linear combinations of basis bivectors.

%Finally, but not least, the geometric product between two vectors (one of the greatest contributions to GA) is defined. 
Any geometric entity can be multiplied by another entity through the geometric product, and the result can be a vector, bivector, trivector, or in general, a multivector.          

\begin{equation}
\bm{a}\bm{b} = \bm{a}\cdot \bm{b}+\bm{a}\wedge \bm{b}
\label{eq:geometricprod}
\end{equation}

The result of geometric multiplication is a linear combination of the inner or scalar product and the wedge product. If the values of vectors $\bm{a}$ and $\bm{b}$ are substituted in (\ref{eq:geometricprod}), we obtain the following:

\begin{equation}
\bm{A}=\bm{a}\bm{b} = \langle\bm{A}\rangle_0+\langle\bm{A}\rangle_2=(\alpha_1\beta_1+\alpha_2\beta_2)+(\alpha_1\beta_2-\alpha_2\beta_1)\bm{\sigma}_1\bm{\sigma}_2
\end{equation}

\noindent where $ \langle\bm{A}\rangle_0$ is the scalar part and $\langle\bm{A}\rangle_2$ is the bivector part.
Noninitiated readers of GA may consult either the final Appendix or the specific references \cite{bernard1989multivectors,dorst2010geometric,hestenes2012clifford}.

\section{Circuit analysis and powers in Geometric Algebra}

The study and analysis of AC electrical circuits in the frequency domain has traditionally been performed using complex numbers algebra as the fundamental analysis tool. Recently, more advanced techniques such as quaternions \cite{talebi2015quaternion,barry2016application,do2018electrical}, have been proposed. However, the promising method for the analysis of electrical systems is the one based on GA because of its inherent robustness, compactness and flexibility in representing the multicomponent nature of the current and voltage  and the associated power flow \cite{menti2007geometric,castro2013use,castro2010use}. The simultaneous handling of harmonics in nonsinusoidal and nonlinear environments has been properly demonstrated in the literature \cite{castilla2008geometric,montoya2018power}. New power terms that comply with the PCoE can be defined using GA. This is possible in none of the power theories formulated so far. One of the most comprehensive GA power theory from a formal and mathematical point of view is the theory formulated by C-N \cite{castro2010use} and extended in his PhD thesis \cite{castro2013use}, which defines new power concepts such as the geometric reactive power and the degraded power. This thesis presents for the first time an unprecedented form of reactive power caused by the interaction of current and voltage harmonics of different frequency. A more comprehensive analysisis is possible and conditions for quasi-optimum compensation of nonactive currents (i.e., currents that do not contribute to the active power $P$ in any situation of voltage and current distortion) can be established. The decomposition of the current into an in-phase term and a quadrature term enables the design of passive compensators that contribute to near-optimal compensation.

\subsection{Castro-Núñez  proposal}

The basics postulated by C-N are defined below. These have been presented in several scientific publications and validated through examples scrutinised over time by other authors.
First, the base transformation is presented \cite{castro2012advantages}, which is used to transform variables defined in the time domain to the geometric domain as follows:

\begin{equation}
\arraycolsep=1.4pt\def\arraystretch{1.4}
\begin{array}{lcl}
\varphi_{c1}(t) = \sqrt{2}\cos{\omega t} \quad &\longleftrightarrow  & \quad  \phantom{-}\bm{\sigma}_1 \\
\varphi_{s1}(t) = \sqrt{2}\sin{\omega t} \quad & \longleftrightarrow & \quad  \bm{-\sigma}_2 \\
\varphi_{c2}(t) = \sqrt{2}\cos 2\omega t \quad & \longleftrightarrow & \quad  \boldsymbol{\sigma}_2\bm{\sigma}_3 \\
\varphi_{s2}(t) = \sqrt{2}\sin 2\omega t \quad & \longleftrightarrow & \quad  \boldsymbol{\sigma}_1\bm{\sigma}_3 \\
& \vdots\\
\varphi_{cn}(t) = \sqrt{2}\cos n\omega t \quad & \longleftrightarrow & \quad  \boldsymbol{\bigwedge\limits_{i=2}^{n+1} \sigma}_i \\
\varphi_{sn}(t) = \sqrt{2}\sin n\omega t \quad & \longleftrightarrow & \quad  \boldsymbol{\bigwedge\limits_{\substack{i=1\\i \ne 2}}^{n+1} \sigma}_i \\
\end{array}
\label{eq:castrotransform1}
\end{equation}

where $\bigwedge_{n} \bm{\sigma}_i$ is the product of $n$ vectors $\bm{\sigma}_i$. The subscript $c$ indicates cosine and the subscript $s$ indicates sine. For example,  a  voltage $v(t)$   

\begin{equation}
v(t)=\sqrt{2}\Big[ 230\cos\omega t+ 20\sin4\omega t\Big] 
\end{equation}

 can be transformed to the geometric domain as:

\begin{equation}
\bm{V}=\underbrace{230\bm{\sigma}_1}_{\langle \bm{u} \rangle_1} + \underbrace{20\bm{\sigma}_{1345}}_{\langle \bm{u} \rangle_4}
\end{equation}

One of the significant contributions  by C-N is the definition of impedance and admittance in the geometric domain. To do this, Kirchhoff's laws are set out in the time domain and then completely transferred to the geometric domain, such that a relation between voltage and current is found. For example, consider the circuit in Figure \ref{fig:inductive_load} and a voltage source  $v(t)=\sqrt{2}V\cos\omega t$. According to  \cite{bernard1989multivectors}, a rotating vector $\bm{n}(t)$  can be represented in $\mathcal{G}_{2}$ domain as follows:

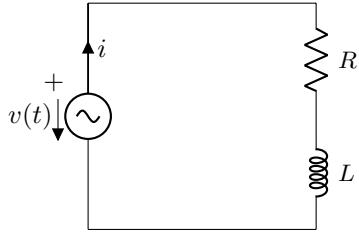
\begin{figure} 
	\centering 
	\begin{circuitikz}[scale=1.5,/tikz/circuitikz/bipoles/length=1cm] \draw
		(0,0) to[sV, v<=$v(t)$, i=$i$] (0,2)  --  (2,2) to[R, l^=\small $R$] (2,1)		
		(2,1)  to [L, cute inductors,  l^=\small $L$] (2,0) -- (0,0) 
		;
		
		\draw (-0.3,1.3) node{$+$};
	\end{circuitikz}
	\caption{Inductive load}
	\label{fig:inductive_load}
\end{figure}

\begin{equation}
	\bm{n}(t)=e^{-\frac{1}{2}\omega t\bm{\sigma}_{12}}\bm{N}e^{\frac{1}{2}\omega t\bm{\sigma}_{12}}=\bm{R^{\dagger}}\bm{N}\bm{R}
	\label{eq:rotor}
\end{equation}

where $\bm{R}$ is a rotor and $\bm{N}$ is a vector or geometric phasor. It's easy to check that $v(t)$ is the projection of $\bm{n}(t)$ over $\bm{\sigma}_1$

\begin{equation}
v(t)= \bm{n}(t)\cdot\bm{\sigma}_1 = \sqrt{2}V\cos(\omega t) 
\end{equation}

This approach is equivalent to that used in the complex domain by means of rotating vectors where $v(t)=Re[\sqrt{2}\bm{\vec{V}}e^{j\omega t}]$ and $\bm{\vec{V}}$ is a complex phasor. If the analysis equation is applied to the proposed circuit as follows:

\begin{equation}
	v(t)=Ri(t)+L\frac{di(t)}{dt}
\end{equation}

and is combined with (\ref{eq:rotor}), the result is:

\begin{equation}
\begin{aligned}
	e^{-\frac{1}{2}\omega t\bm{\sigma}_{12}}\bm{V}e^{\frac{1}{2}\omega t\bm{\sigma}_{12}} = Re^{-\frac{1}{2}\omega t\bm{\sigma}_{12}}\bm{I}e^{\frac{1}{2}\omega t\bm{\sigma}_{12}}+ L\frac{d(e^{-\frac{1}{2}\omega t\bm{\sigma}_{12}}\bm{I}e^{\frac{1}{2}\omega t\bm{\sigma}_{12}})}{dt}
\end{aligned}
\end{equation}

In the previous equation, we must first perform the derivative and choose a specific instant for the steady state, for example $t=0$, so we get

\begin{equation}
	\bm{V}=\bm{I}R+\bm{I}\cdot L\omega \bm{\sigma}_{12}
	\label{eq:ecuacion1}
\end{equation} 

Observe in (\ref{eq:ecuacion1}) the scalar product between the current $\bm{I}$ and the term $L\omega \bm{\sigma}_{12}$. If the terms are reordered, the result is the geometric impedance  as follows: 

\begin{equation}
	\bm{Z}=\bm{V}\bm{I}^{-1}=R-L\omega\bm{\sigma}_{12}=R+X_L\bm{\sigma}_{12}
	\label{eq:impedance_inductive}
\end{equation}

\noindent where $X_L$ is the inductive geometric  reactance. It should be noted that $\frac{\bm{V}}{\bm{I}}$ is ambiguous in GA, so it should  be avoided. Instead, left or right multiplication by the inverse is the proper choice. Similarly, a geometric capacitive reactance can be obtained for a circuit with capacitors. In equation (\ref{eq:impedance_inductive}), the inverse of the current was multiplied from the right, which results in a negative geometric reactance. If the multiplication is performed from the left, then a positive reactance is obtained, although the C-N prefers the first method. Nevertheless, in terms of practical results, the choice of method has a negligible effect \cite{castro2019theorems}.

Admittance is defined as follows:

\begin{equation}
	\bm{Y}=\bm{Z}^{-1}=\frac{\bm{Z}^{\dagger}}{\bm{Z}^{\dagger}\bm{Z}}=\frac{\bm{Z}^{\dagger}}{\|\bm{Z}\|^2}=G+B\bm{\sigma}_{12}
\end{equation}

In general, the impedance of a load at the frequency of harmonic $n$ is as follows:

\begin{equation}
	\bm{Z}_n=R+\left(\frac{1}{n\omega C}-n\omega L\right)\bm{\sigma}_{12}
\end{equation}

Therefore, any nonsinusoidal voltage, such as the following:

\begin{equation}
\begin{aligned}
v(t)=  \sum_{i=1}^{n}v_i(t) =& D_1\cos(\omega t)+ E_1\sin(\omega t) + \\
&+\sum_{h=2}^{d}D_h\cos(h\omega t) + \sum_{h=2}^{k}E_h\sin(h\omega t) 
\label{eq:voltage_gen}
\end{aligned}	
\end{equation}

\noindent can be transformed to the geometric domain as the following:

\begin{gather}
\bm{V}=D_1\bm{\sigma}_1-E_1\bm{\sigma}_2+ \sum_{h=2}^{d}\left[D_h\bigwedge_{i=2}^{h+1}\bm{\sigma}_i\right]+ \sum_{h=2}^{k}\left[E_h\bigwedge_{i=1,i\neq2}^{h+1}\bm{\sigma}_i\right] 
\label{eq:voltgeononsen}
\end{gather} 

Similarly, the resulting current for any load will be as follows:  

\begin{flalign}
\begin{split}
\bm{I} &=G_1D_1\bm{\sigma}_1-G_1E_1\bm{\sigma}_2 +\sum_{h=2}^{d}\left[G_hD_h\bigwedge_{i=2}^{h+1}\bm{\sigma}_i\right]+\sum_{h=2}^{k}\left[G_hE_h\bigwedge_{i=1,i\neq2}^{h+1}\bm{\sigma}_i\right]- \\
&-B_1E_1\bm{\sigma}_1-B_1D_1\bm{\sigma}_2+ \sum_{h=2}^{d}\left[B_hD_h\bigwedge_{i=1,i\neq 2}^{h+1}\bm{\sigma}_i\right] -\sum_{h=2}^{k}\left[B_hE_h\bigwedge_{i=2}^{h+1}\bm{\sigma}_i\right] 
\end{split}
\label{eq:currentDescomp}
\end{flalign}

\noindent  For each harmonic, the admittance is $\bm{Y_n}=G_n+B_n\bm{\sigma}_{12}$. In equation (\ref{eq:currentDescomp}), the current has been decomposed into the following two components: 

	\begin{equation}
\bm{I}=\bm{I}_{||}+\bm{I}_{\perp}=\bm{I}_g+\bm{I}_b
\label{eq:currentDescomp_lite}
\end{equation}

\noindent where $\bm{I}_g$ is the in-phase current caused by the conductance $G_n$ of each harmonic and $\bm{I}_b$ is the quadrature current caused by the susceptance $B_n$. Note that this decomposition resembles that of Shepherd and Zakikhani.

Once the voltage and current are obtained, the geometric apparent power is defined as the product of both magnitudes as follows:

\begin{align}
\begin{split}
\bm{M}  & =  \bm{V}\bm{I} = \underbrace{\overbrace{\vphantom{\langle \bm{M}_g\rangle_0 + \sum_{i=1}^{n+1}\langle \bm{M}_g\rangle_i}\langle \bm{M}_g\rangle_0}^{P} + \overbrace{\sum_{i=1}^{n+1}\langle \bm{M}_g\rangle_i}^{\bm{CN}_d}}_{\bm{M}_g}
+ \underbrace{\vphantom{\langle \bm{M}_g\rangle_0 + \sum_{i=1}^{n+1}\langle \bm{M}_g\rangle_i} \bm{CN}_{r(ps)} + \bm{CN}_{r(hi)}}_{\bm{M}_b = \bm{CN}_r } 
\end{split}
\label{eq:Mpower}
\end{align}

\noindent where 

\begin{flalign*}
\qquad \bm{M}_g & \text{ is the parallel geometric apparent power }  &\\
\bm{M}_b & \text{ is the quadrature geometric apparent power }  &\\
\bm{P}  & \text{ is the active power }  &\\
\bm{CN}_d  & \text{ is the degraded power}  &\\
\bm{CN}_r  & \text{ is the quadrature geometric power or reactive} \\ &\text{geometric power}  \\
\bm{CN}_{r(ps)}  & \text{ is the reactive geometric  power due to voltage} \\ & \text{and current  phase shift of same components} \\
\bm{CN}_{r(hi)}  & \text{ is the reactive geometric  power due to voltage} \\ & \text{and current cross products} &\\
\end{flalign*}

In sinusoidal conditions, \cite{castro2012ieee,castro2012advantages} demonstrates that equation (\ref{eq:Mpower}) is reduced to the well-known equation $\bm{S}=P+jQ$, since $\bm{CN}_{r(hi)}=\bm{CN}_d=0$, and $\bm{CN}_{r(ps)} = Q\bm{\sigma}_{12}$. Additionally, C-N demonstrated the following properties of $\bm{M}$:

\begin{itemize}[topsep=8pt,itemsep=4pt,partopsep=4pt, parsep=4pt]
		\item $\|\bm{M}\|\ne \|\bm{V}\|\|\bm{I}\|$
		\item $\|\bm{M}\|=\sqrt{P^2+CN_{r(ps)}^2+CN_{r(hi)}^2+CN_{d}^2}$
		\item $\bm{M}$ is a conservative quantity that takes into account the net direction of power flows in the branches of any circuit. The conservation of energy principle and Tellegen's Theorem \cite{castro2019theorems} are both satisfied.
\end{itemize} 

Surprisingly, C-N had to include a correction factor to adjust the balance of active power as follows:      

\begin{equation}
	f=(-1)^{\frac{k(k-1)}{2}}
\end{equation}

Otherwise, $P=\sum P_i$ is not always satisfied because not all the values of $k$-vectors square to $+1$. This is one of the weaknesses of his theory. Additionally, it does not include the possibility  to handle the presence of sub- and inter-harmonics. Finally, the decomposition of currents into an in-phase component $\bm{I}_g$ and a quadrature component $\bm{I}_b$ has a limited scope since it does not provide information for the minimum current necessary to deliver the same power $P$ of the original circuit. Therefore, only a partial compensation can be performed through passive elements.              

\section{New proposal and  methodology}

%Starting from advanced systems with non-sinusoidal and nonlinear sources to simpler circuits made up of linear loads and sinusoidal sources, the different alternatives have failed to provide an accurate and detailed explanation about how energy flows or  an interpretation in practical engineering terms.

To address the above-mentioned flaws found in C-N theory, a new reformulation is proposed with major upgrades that reveal some baseline issues. The first and main correction is related to the definition of geometric apparent power $\bm{M}=\bm{V}\bm{I}$. We propose a slightly but different new definition:

\begin{equation}
	\bm{M}=\bm{V}\bm{I}^{\dagger}
\end{equation}

This is important and necessary because if the reverse of the current is not performed, the result is: a) having to include $f=(-1)^{\frac{k(k-1)}{2}}$, an unnatural corrective term to calculate active power $P$ and b) miscalculating the rest of nonactive power terms. In fact, this is one of the main drawbacks of C-N theory.

The above can be verified with a very simple example by supposing that a voltage at a fundamental frequency $\bm{V}_1=\alpha_1\bm{\sigma}_1+\alpha_2\bm{\sigma}_2$ (for simplicity, we consider $\omega=1$) is applied to a linear load, so a current $\bm{I}_1=\beta_1\bm{\sigma}_1+\beta_2\bm{\sigma}_2$ is obtained. The product of both quantities is as follows:

\begin{equation}
\bm{M}_1=	\bm{V}_1\bm{I}_1=\underbrace{(\alpha_1\beta_1+\alpha_2\beta_2)}_{P}+\underbrace{(\alpha_1\beta_2-\alpha_2\beta_1)}_{Q}\bm{\sigma}_{12}
	\label{eq:generic_power}
\end{equation}
which matches  $\bm{V}_1\bm{I}_1^{\dagger}$  because the reverse of a vector is the vector itself. However, supposing the same voltage at twice the frequency used above, i.e., $\omega=2$, results in the following: 
\begin{equation}
\begin{aligned}
	\bm{V}_2&=\alpha_1\bm{\sigma}_{13}+\alpha_2\bm{\sigma}_{23} \\
	\bm{I}_2&=\beta'_1\bm{\sigma}_{13}+\beta'_2\bm{\sigma}_{23}
\end{aligned}
\end{equation}

Remember that the impedance/admitance is different for $\omega=2$, so the coefficients $\beta_1$ and $\beta_2$ changes to $\beta'_1$ and $\beta'_2$, respectively. Consequently, the new power expression is: 

\begin{equation}
\begin{aligned}
		\bm{M}_2=\bm{V}_2\bm{I}_2=-(\alpha_1\beta'_1+\alpha_2\beta'_2)+(-\alpha_1\beta'_2+\alpha_2\beta'_1)\bm{\sigma}_{12}
\end{aligned}
\end{equation}

which is clearly different from expression (\ref{eq:generic_power}). Moreover, after the correction  factor $f$ proposed by C-N is applied, the value of the geometric reactive/quadrature power remains different and thus incorrect. However, if we apply the proposed definition:

\begin{equation}
\begin{aligned}
	\bm{M}_{2_{new}}=\bm{V}_2\bm{I}_2^{\dagger}=(\alpha_1\beta'_1+\alpha_2\beta'_2)+(\alpha_1\beta'_2-\alpha_2\beta'_1)\bm{\sigma}_{12}
	\end{aligned}
\end{equation}

now it really agrees with (\ref{eq:generic_power}). It is therefore evident, the necessity of performing the reverse of the current to obtain the right value for the geometric apparent power.   

The second main contribution  is the addtion of interharmonics in the transformation from time domain to the geometric domain \cite{MONTOYA2019486}. In addition to (\ref{eq:castrotransform1}),  non-integer multiples of fundamental (interharmonics) are also defined

\begin{equation}
\arraycolsep=1.4pt\def\arraystretch{2.2}
\begin{array}{ll}
X_{c_n p_k}  =\left({\bigwedge\limits_{i=2}^{n+1} \bm{\sigma}_i}\right)\bm{\sigma}_{(k+n+2)} \\
X_{s_n p_k}  =\left({\bigwedge\limits_{\substack{i=1\\i \ne 2}}^{n+1} \bm{\sigma}_i}\right)\bm{\sigma}_{(k+n+2)}
\end{array}
\end{equation}

where $p_k$ is the interharmonic $k$ that exists between harmonic $n$ and harmonic $n+1$. 

The third main contribution is related to the decomposition of the total current and its allocation to physical phenomena that have economic and engineering relevance. The original decomposition of C-N into in-phase current and quadrature current is performed according to equation (\ref{eq:currentDescomp_lite}) as follows:

\begin{equation}
\begin{aligned}
	\bm{I}_g&=\sum_{n}G_n\langle \bm{V}\rangle_n \\
	\bm{I}_b&=\sum_{n}B_n\bm{\sigma}_{12}\langle \bm{V}\rangle_n
	\end{aligned}
\end{equation}

However, this decomposition can be expanded to include other interesting terms that have been described in the scientific literature: active or Fryze’s current $\bm{I}_a$, which is the minimum current necessary to obtain the active power $P$ of the load. Moreover, the current that has frequency components that are not present in the supply, $\bm{I}_G$, can also be defined. The active current is obtained using the concept of Fryze’s equivalent load, i.e., a load equivalent conductance $G_e$ that demands the same power $P$ as the original load when the same voltage $v(t)$ (with RMS $\|\bm{V}\|$)  is applied as follows:       

\begin{equation}
	G_e=\frac{P}{\|\bm{V}\|^2}
\end{equation}

The active current can then be defined as:  

\begin{equation}
	\bm{I}_a=G_e\bm{V}
	\label{eq:ecuacion2}
\end{equation}

The current defined in (\ref{eq:ecuacion2}) is already included in  $\bm{I}_g$, and therefore, it can be inferred that there is another additional current component as follows: 

\begin{equation}
	\bm{I}_s=\bm{I}_g-\bm{I}_a
\end{equation}

which coincides with the scattered current defined in the CPC theory proposed by Czarnecki. The term scattered current will be used for $\bm{I}_s$ to avoid the introduction of new terms in addition to those already used in the existing literature. Note that $\bm{I}_s$ can contain all harmonic terms at once, in contrast with complex numbers algegra, where this is not possible. 
Once more, GA demonstrates its potential by naturally describing the basic components of electrical interest. The complete decomposition of the current then would be as follows:

\begin{equation}
	\bm{I}=\underbrace{\bm{I}_a+\bm{I}_s}_{\bm{I}_g}+\bm{I}_b+\bm{I}_G
	\label{eq:current_GAPoT}
\end{equation}

\noindent with the following:

\begin{equation*}
\begin{array}{ll}
\bm{I}_a: & \text{minimun current for active power } P \text{ in the load} \\
\bm{I}_s: & \text{current due to changes in conductance  with frequency} \\
\bm{I}_b: & \text{quadrature current with voltage} \\
\bm{I}_G: & \text{harmonic current generated  by the load} \\
\end{array}
\end{equation*}

It can be readly demonstrated (not performed due to the extension os this work) that the four components of the current are orthogonal, in addition to the already well-known quadrature between $\bm{I}_g$ and $\bm{I}_b$ (\ref{eq:currentDescomp}) as follows:

\begin{equation}
\begin{aligned}
	\bm{I}_a \cdot \bm{I}_s=0 \\
\bm{I}_a \cdot \bm{I}_b=0 \\
\bm{I}_s \cdot \bm{I}_b=0 \\
\bm{I}_a \cdot \bm{I}_G=0 \\
\bm{I}_s \cdot \bm{I}_G=0 \\
\bm{I}_b \cdot \bm{I}_G=0 \\
\end{aligned}
\end{equation}

and therefore, the following is also satisfied:   

\begin{equation}
	\|\bm{I}\|^2=\|\bm{I}_a\|^2+\|\bm{I}_s\|^2+\|\bm{I}_b\|^2+\|\bm{I}_G\|^2
	\label{eq:current_cuadratico}
\end{equation}

Once the decomposition of currents is presented, note that  left multiplying the reverse of equation (\ref{eq:current_GAPoT}) by $\bm{V}$, the geometric power equation is derived: 

\begin{equation}
	\bm{M}=\bm{V}\bm{I}^\dagger=\bm{M}_a+\bm{M}_s+\bm{M}_b+\bm{M}_G
\end{equation}

which is totally unrelated to the following traditional apparent power formula:

\begin{equation}
	\|\bm{V}\|^2\|\bm{I}\|^2=\|\bm{V}\|^2\|\bm{I}_a\|^2+\|\bm{V}\|^2\|\bm{I}_s\|^2+\|\bm{V}\|^2\|\bm{I}_b\|^2+\|\bm{V}\|^2\|\bm{I}_G\|^2
\end{equation}

 where the expression (\ref{eq:current_cuadratico}) multiplied by $\|\bm{V}\|^2$ is used.% 

Finally, the power factor is defined based on the geometric apparent power as
\begin{equation}
pf= \frac{P}{\|\bm{M}\|}
\end{equation}

\section{Case of study}
Different circuits will be solved to validate the proposed power theory. Additionally, the results will be compared with other theories to demonstrate  its superiority. The Matlab environment has been chosen to facilitate the resolution of the proposed circuits. Specifically, the Clifford Multivector Toolbox  developed by Sangwine and Hitzer \cite{sangwine2017clifford} has been used.      
\subsection{Sinusoidal case}
The first example is a linear circuit supplied by a sinusoidal source. This is very basic goal, but frequently used in various disciplines to demonstrate that the proposed theory works with both complex cases and the most basic ones. Only when it is demonstrated that the simplest circuits can be solved with a new proposed theory, is it possible to move on to larger and more complex circuits.

\begin{figure}
	\centering
	\includegraphics[width=0.5\textwidth]{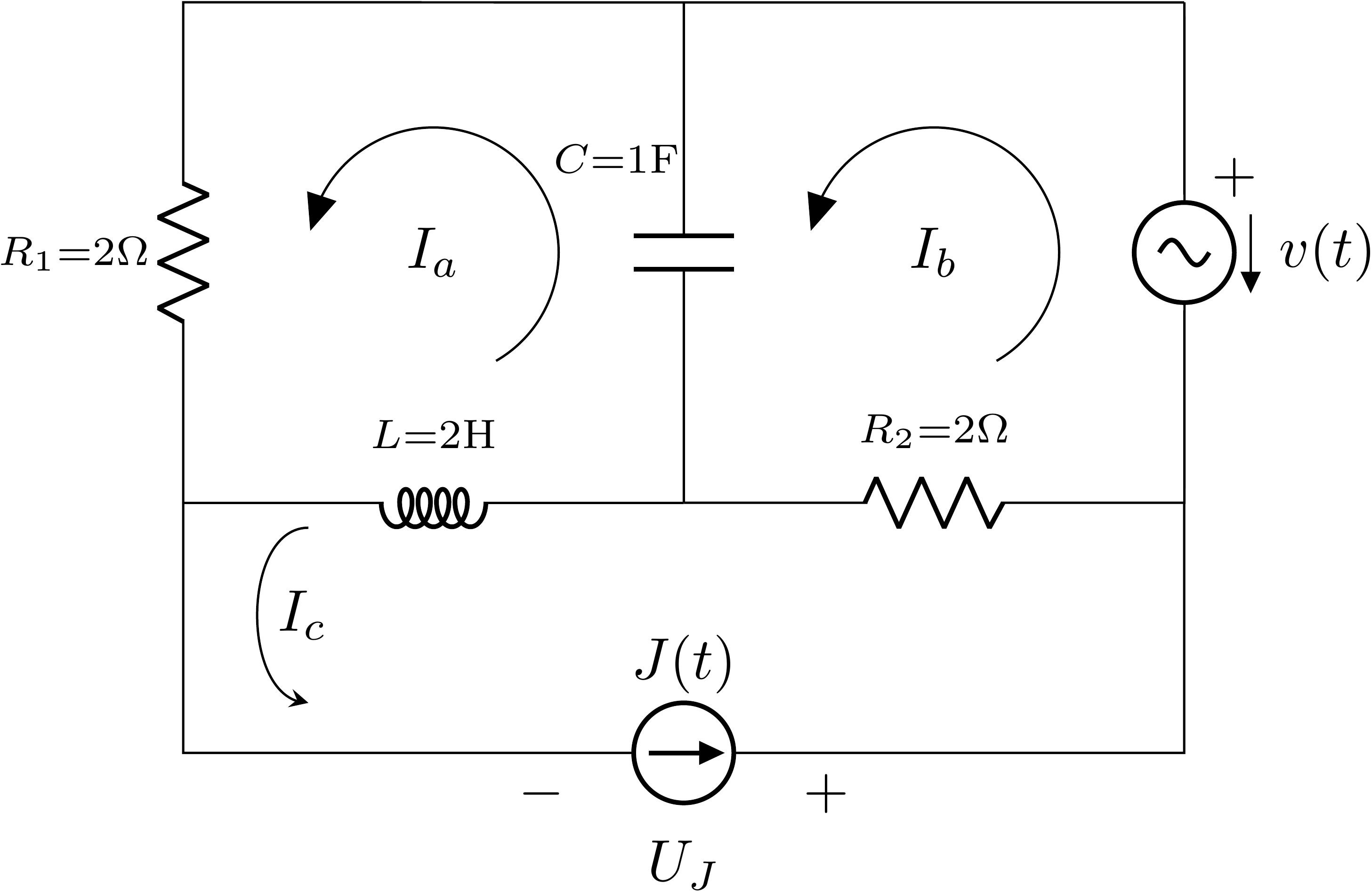}
	\caption{RLC circuit with sinusoidal sources}
	\label{fig:RLC_sinusoidal}
\end{figure}

The circuit in Figure \ref{fig:RLC_sinusoidal} represents a system with linear RLC elements, an ideal voltage source $v(t)=50\sqrt{2}\cos\omega t$ and an ideal current source $J(t)=20\sqrt{2}\cos(\omega t + 90)$. We first start by solving it using KCL/KVL (mesh current method, for example) and phasors in complex algebra:

\begin{equation}
\left(	\begin{array}{ccc}
2+j & j & 0 \\
j & 2-j & 0 \\
-2j & -2 & -1 \\			
\end{array} \right)
\left( \begin{array}{c}
\vec{I}_a\\
\vec{I}_b\\
\vec{U}_J
\end{array} \right)
= \left( \begin{array}{c}
-40\\
50+40j\\
40-40j
\end{array} \right)
\end{equation}

If the unknown vector is solved, the result is as follows:

\begin{equation}
\left( \begin{array}{c}
\vec{I}_a\\
\vec{I}_b\\
\vec{U}_J
\end{array} \right)
= \left( \begin{array}{r}
-6.66-1.66j\\
10+28.33j\\
-63.33-3.33j
\end{array} \right)
\end{equation}

In Table \ref{tab:RLC_resumen}, the full results for the current, the voltage and the apparent power are shown. The data in the table show how the balance of complex apparent powers is achieved with sinusoidal sources as expected.

\newcolumntype{L}{>{$}l<{$}} % math-mode version of "l" column type
\newcolumntype{R}{>{$}r<{$}} % math-mode version of "r" column type
\begin{table}[]
	\centering
	%	\resizebox{0.5\textwidth}{!}{
	\begin{tabular}{@{}lRRRl@{}}
		\toprule
		& \multicolumn{1}{c}{$\vec{I}$} & \multicolumn{1}{c}{$\vec{V}$} & \multicolumn{1}{c}{$\vec{S}$} & \\ \cmidrule(l){2-2} \cmidrule(l){3-3} \cmidrule(l){4-4}
		$\bm{R_1}$ & -6.67-1.67j           & -13.33-3.33j          & 94.47              &   \\
		$\bm{R_2}$ & 10.00+8.33j              & 20.00+16.67j             & 388.86    &            \\
		$\bm{C}$   & 16.67+30.00j             & 30.00-16.67j             & -1177.90j&              \\
		$\bm{L}$   & -6.67-21.67j          & 43.33-13.33j          & 1027.90j    &           \\ \cmidrule{4-4}
		&                       &                  \text{(Demanded)}     & \bm{433.33-150.00j}   &        \\ \midrule
		$\bm{V}$   & 10.00+28.33j             & 50.00                    & 500.00-1416.50j           \\
		$\bm{J}$   & 20.00j                   & -63.33-3.33j          & -66.66+1266.50j     &    \\ \cmidrule{4-4}
		&                       &               \text{(Generated)}          & \bm{433.33-150.00j}      &   \\ \bottomrule
	\end{tabular}
%}
	\caption{Summary table for RLC circuit (phasor solution).}
	\label{tab:RLC_resumen}
\end{table}

If the same circuit is solved using GA, applying Kirchhoff's laws results in the following:  

 \begin{equation}
 %  \resizebox{\hsize}{!}{
 %  	$
 \left(	\begin{array}{ccc}
 2-\bm{\sigma}_{12} & -\bm{\sigma}_{12} & 0 \\
 -\bm{\sigma}_{12} & 2+\bm{\sigma}_{12} & 0 \\
 2\bm{\sigma}_{12} & -2 & -1 \\			
 \end{array} \right)
 \left( \begin{array}{c}
 \bm{I}_a\\
 \bm{I}_b\\
 \bm{U_J}
 \end{array} \right)
 = \left( \begin{array}{c}
 -40\bm{\sigma}_{1}\\
 50\bm{\sigma}_{1}+40\bm{\sigma}_{2}\\
 40\bm{\sigma}_{1}-40\bm{\sigma}_{2}
 \end{array} \right) 
 %$}
 \end{equation}

Solving again for the unkown vector, the result is as follows:   
 
 \begin{equation}
%    \resizebox{\hsize}{!}{
% 	$
 \begin{split}
  \left( \begin{array}{c}
 \bm{I}_a\\
 \bm{I}_b\\
 \bm{U}_J
 \end{array} \right)
 &= \left(	\begin{array}{ccc}
 2-\bm{\sigma}_{12} & -\bm{\sigma}_{12} & 0 \\
 -\bm{\sigma}_{12} & 2+\bm{\sigma}_{12} & 0 \\
 2\bm{\sigma}_{12} & -2 & -1 \\			
 \end{array} \right)^{-1} 
 \left( \begin{array}{c}
 -40\bm{\sigma}_{1}\\
 50\bm{\sigma}_{1}+40\bm{\sigma}_{2}\\
 40\bm{\sigma}_{1}-40\bm{\sigma}_{2}
 \end{array} \right)=\\
 &=
 \left( \begin{array}{r}
 -6.66\bm{\sigma}_{1}-1.66\bm{\sigma}_{2}\\
 10\bm{\sigma}_{1}+28.33\bm{\sigma}_{2}\\
 -63.33\bm{\sigma}_{1}-3.33\bm{\sigma}_{2}
 \end{array} \right)
 \end{split}
 %$}
 \end{equation}
 
 Table \ref{tab:RLC_GA_resumen} shows the same results as Table \ref{tab:RLC_resumen} but transferred to the geometric domain. Compliance with PCoE is observed for the geometric power $\bm{M}$.

\begin{table*}[]
	\centering
	\begin{tabular}{@{}lRRRl@{}}
		\toprule
		& \multicolumn{1}{c}{$\bm{I}$} & \multicolumn{1}{c}{$\bm{V}$} & \multicolumn{1}{c}{$\bm{M}$} & \\ \cmidrule(l){2-2} \cmidrule(l){3-3} \cmidrule(l){4-4}
		${R}_1$ & -6.67\bm{\sigma}_{1}-1.67\bm{\sigma}_{2}           & -13.33\bm{\sigma}_{1}-3.33\bm{\sigma}_{2}          & 94.47              &   \\
		${R}_2$ & 10.00\bm{\sigma}_{1}+8.33\bm{\sigma}_{2}              & 20.00\bm{\sigma}_{1}+16.67\bm{\sigma}_{2}             & 388.86    &            \\
		${C}$   & 16.67\bm{\sigma}_{1}+30.00\bm{\sigma}_{2}             & 30.00\bm{\sigma}_{1}-16.67\bm{\sigma}_{2}             & 1177.90\bm{\sigma}_{2}&              \\
		${L}$   & -6.67\bm{\sigma}_{1}-21.67\bm{\sigma}_{2}          & 43.33\bm{\sigma}_{1}-13.33\bm{\sigma}_{2}          & -1027.90\bm{\sigma}_{12}    &           \\ \cmidrule{4-4}
		&                       &              \text{(Demanded)}         & \bm{433.33+150.00\sigma_{12}}   &        \\ \midrule
		$\bm{V}$   & 10.00\bm{\sigma}_{1}+28.33\bm{\sigma}_{2}             & 50.00\bm{\sigma}_{1}                    & 500.00+1416.50\bm{\sigma}_{12}           \\
		$\bm{J}$   & 20.00\bm{\sigma}_{2}                   & -63.33\bm{\sigma}_{1}-3.33\bm{\sigma}_{2}          & -66.66-1266.50\bm{\sigma}_{12}     &    \\ \cmidrule{4-4}
		&                       &            \text{(Generated)}              & \bm{433.33+150.00\sigma_{12}}      &  \\ \bottomrule
	\end{tabular}
	\caption{Summary table for RLC circuit (GA solution).}
	\label{tab:RLC_GA_resumen}
\end{table*}

\subsection{Non sinusoidal case}
One of the major problems with existing power theories is their inability to properly handle nonsinusoidal systems because they cannot verify the principle of conservation of energy  or Tellegen's theorem for the apparent power $S$. The use of GA overcomes this problem  by defining a conservative quantity,  the geometric apparent power $\bm{M}$. 
To illustrate this point, the circuit in Figure \ref{fig:inductive_load} is considered again with $R=1\Omega$, $L=1$H, but with a nonsinusoidal voltage $v(t)=\sqrt{2}[100\sin(\omega t-45)+30\sin(2\omega t + 30)]$. The transformed voltage and current  to the geometric domain is as follows:            

\begin{equation}
\begin{aligned}
\bm{V}&=-70.71\bm{\sigma}_1-70.71\bm{\sigma}_2+25.98\bm{\sigma}_{13}+15.00\bm{\sigma}_{23} \\
\bm{I}&=-42.43\bm{\sigma}_1+14.14\bm{\sigma}_2+5.06\bm{\sigma}_{13}-5.23\bm{\sigma}_{23}
\end{aligned}
\end{equation}

The power obtained taking into account the reverse of the current is as follows:

\begin{equation}
%    \resizebox{\hsize}{!}{
%	$
\bm{M}= \bm{VI}^{\dagger}=2052.9+877.90\bm{\sigma}_3-4211.76\bm{\sigma}_{12}-1731.32\bm{\sigma}_{123}  
%$}
\end{equation}

\begin{equation}
 \|\bm{M}\|= 5071.7
\end{equation}

In Table \ref{tab:power_RL}, a more detailed analysis of the power balance is shown. To stress the power concept defined by C-N against the proposed method, the term $\bm{M}_{CN}=\bm{V}\bm{I}$ is also included. The data show how our proposal properly adds the contribution of the first and second harmonic ($\bm{M}_1$ and $\bm{M}_2$), while C-N fails to subtract the two terms. This is a clear indication that power must be defined using the reversed current.

\definecolor{LightCyan}{rgb}{1,0.5,0.5}
\begin{table}[]
	\centering
	\begin{tabular}{@{}lrrrrr@{}}
		\toprule
		&  \multicolumn{4}{c}{\textbf{k-vector}} \\ \cmidrule{3-6}
		& \multicolumn{1}{c}{$\|\cdot\|$} & \multicolumn{1}{c}{$\bm{\sigma}_0$} & \multicolumn{1}{c}{$\bm{\sigma}_3$} &  \multicolumn{1}{c}{$\bm{\sigma}_{12}$}  & \multicolumn{1}{c}{$\bm{\sigma}_{123}$} \\ \midrule
	    $\bm{M}_{CN}$    & 4410.90                   & 2052.90                 &         902.37               & -3788.23                & -276.31      \\ 
		$\bm{M}_{GA}$    & 5071.70                   & 2052.90                 &         877.90               & -4211.70                & -1731.32                   \\ 
		$\bm{M}_{1}$    & 4669.00                   & 2000.00                 &         890.13               & -4000.00                & -1003.82 \\
		$\bm{M}_{2}$    & 759.64                   & 52.90                 &         -12.23               & -211.70                & -727.50  \\ [0.1cm] \hdashline[1.5pt/5pt] 
		$\bm{M}_{R}$    & 2313.50                   & 2052.90                 &         577.10               &                 &  \\
		$\bm{M}_{L}$    & 4563.70                   &                  &         300.80               & -4211.70                & -1731.32   \\ \bottomrule
	\end{tabular}
	\caption{Power decomposition for circuit in Figure \ref{fig:inductive_load}.}
	\label{tab:power_RL}
\end{table}

In Table \ref{tab:currents_non_sinusoidal}, the analysis of currents considering the decomposition described in the equation (\ref{eq:current_GAPoT}) is shown. It can be verified that the sum of the active current $\bm{I}_a$ plus the scattered current $\bm{I}_s$ results in the in-phase current $\bm{I}_g$. Then, add the quadrature or reactive current $\bm{I}_b$ to obtain the total current $\bm{I}$. It can also be shown that all components are in quadrature. GA makes it possible to simultaneously solve the system for all currents and all harmonic components, demonstrating that the principle of superposition is embedded in GA itself.

\begin{table}
	\centering
	\begin{tabular}{@{}lrrrrr@{}}
		\toprule
		& \multicolumn{5}{c}{k-vector} \\ \cmidrule{2-6}
		& \multicolumn{1}{c}{$\bm{\sigma}_1$} & \multicolumn{1}{c}{$\bm{\sigma}_2$} & \multicolumn{1}{c}{$\bm{\sigma}_{13}$} & \multicolumn{1}{c}{$\bm{\sigma}_{23}$} & \multicolumn{1}{r}{$\|\cdot\|$} \\ \cmidrule(l){2-5} \cmidrule(l){6-6}
		$\bm{I}_s$ & -0.82       & -0.82      & -3.36    & -1.94        & 4.06 \\
		$\bm{I}_a$ & -13.32       & -13.32      & 4.89    & 2.82        & 19.66 \\ \cmidrule{2-6}
		$\bm{I}_g$ & -14.14       & -14.14      & 1.53    & 0.88        & 20.08 \\
		$\bm{I}_b$ & -28.28       & 28.28       & 3.53    & -6.11       & 40.62  \\ \cmidrule{2-6}
		$\bm{I}$ & \textbf{-42.42}   & \textbf{14.14}   & \textbf{5.06}   & \textbf{-5.23}   & \textbf{45.31}\\ \bottomrule
	\end{tabular}
	\caption{Current summary  for non sinusoidal case (in Amperes).}
	\label{tab:currents_non_sinusoidal}
\end{table}

For comparison, the same circuit can be solved with a more distorted voltage, for example, with three harmonics as follows:

\begin{equation}
	v(t)=100\sqrt{2}\sin(\omega t - 30)+50\sqrt{2}\sin(2\omega t + 45)	+ 10\sqrt{2}\sin(3\omega t + 75)
\end{equation}

The transformation of the voltage and current (after applying the generalised Ohm's law) to the geometric domain is as follows: 

\begin{equation}
\begin{aligned}
\bm{V}&=-50.00\bm{\sigma}_1-86.60\bm{\sigma}_2+35.35\bm{\sigma}_{13}+35.35\bm{\sigma}_{23}+2.59\bm{\sigma}_{134}+9.66\bm{\sigma}_{234}\\
\bm{I}&=-44.64\bm{\sigma}_1+2.68\bm{\sigma}_2+10.40\bm{\sigma}_{13}-6.24\bm{\sigma}_{23}+1.64\bm{\sigma}_{134}+0.16\bm{\sigma}_{234}
\end{aligned}
\end{equation}

Tables \ref{tab:currents_non_sinusoidal_3harm} and \ref{tab:power_non_sinusoidal_3harm} are obtained again with the decomposition of the current and power. The data demonstrate once more how the power balance is satisfied and, therefore, the PCoE is also satisfied. The multicomponent quality of $\bm{M}$ occurs naturally from the application of GA. In addition, a decomposition based on the engineering usefulness of the current components is achieved.

\begin{table*}
	\centering
	\begin{tabular}{@{}lrrrrrrr@{}}
		\toprule
		& \multicolumn{7}{c}{k-vector} \\ \cmidrule{2-8}
		& \multicolumn{1}{c}{$\bm{\sigma}_1$} & \multicolumn{1}{c}{$\bm{\sigma}_2$} & \multicolumn{1}{c}{$\bm{\sigma}_{13}$} & \multicolumn{1}{c}{$\bm{\sigma}_{23}$} & \multicolumn{1}{c}{$\bm{\sigma}_{134}$} & \multicolumn{1}{c}{$\bm{\sigma}_{234}$} & \multicolumn{1}{r}{$\|\cdot\|$} \\ \cmidrule(l){2-7} \cmidrule(l){8-8}
		$\bm{I}_s$ & -1.4692  &  -2.5447  &  -3.9525  &  -3.9525  &  -0.3716  &  -1.3870 & 6.4761\\
		$\bm{I}_a$ & -8.5308  &  -14.7758  &  6.0322  &  6.0322  &  0.4416  &  1.6480 & 19.1516\\  \cmidrule{2-8}
		$\bm{I}_g$ & -10.0000  &  -17.3205  &  2.0797  &  2.0797  &  0.0700  &  0.2611 & 20.2169\\
		$\bm{I}_b$ & -34.6410  &  20.0000  &  8.3189  &  -8.3189  &  1.5664  &  -0.4197 & 41.7257\\ \cmidrule{2-8}
		$\bm{I}$ & \textbf{-44.6410}   & \textbf{2.6795}   & \textbf{10.3986}   & \textbf{-6.2392}   & \textbf{1.6363} &  \textbf{-0.1586} &  \textbf{46.3655}\\ \bottomrule
	\end{tabular}
	\caption{Current summary in amperes for non sinusoidal case (3 harmonics)}
	\label{tab:currents_non_sinusoidal_3harm}
\end{table*}

\begin{table*}
	\centering
		\resizebox{\textwidth}{!}{
	\begin{tabular}{@{}lrrrrrrrrrr@{}}
		\toprule
		& \multicolumn{9}{c}{k-vector} \\ \cmidrule{2-10}
		& \multicolumn{1}{c}{$\bm{\sigma}_0$} & \multicolumn{1}{c}{$\bm{\sigma}_3$} & \multicolumn{1}{c}{$\bm{\sigma}_{4}$} & \multicolumn{1}{c}{$\bm{\sigma}_{12}$} & \multicolumn{1}{c}{$\bm{\sigma}_{34}$} & \multicolumn{1}{c}{$\bm{\sigma}_{123}$} & \multicolumn{1}{c}{$\bm{\sigma}_{124}$} & \multicolumn{1}{c}{$\bm{\sigma}_{1234}$} & \multicolumn{1}{r}{$\|\cdot\|$} \\ \cmidrule(l){2-9} \cmidrule(l){10-10}
		$\bm{M}_s$ &  & -398.0023 & -110.5834 &  & -167.0791 & 182.6958 & -7.9491 & -29.5584 &  \\
		$\bm{M}_a$ & 2149.7615 & 1648.0240 & 147.7578 &  &  & & & 88.3173 &\\ \cmidrule{2-9}
		$\bm{M}_g$ & 2149.7615 & 1250.0217 & 37.1744 &  & -167.0791 & 182.6958 & -7.9491 & 58.7589 &\\
		$\bm{M}_b$ &  & 213.1451 & -18.2830 & -4604.4515 & 145.4983 & -3068.2350 & -172.1036 & 229.7337 &\\ \cmidrule{2-10}
		$\bm{M}$ & \textbf{2149.7615}   & \textbf{1463.1668}   & \textbf{18.8914}   & \textbf{-4604.4515}   & \textbf{-21.5808} &  \textbf{-2885.5392} & \textbf{-180.0527} & \textbf{288.4926} &  \textbf{6033.70}\\ \bottomrule
	\end{tabular}
}
	\caption{Power summary  for non sinusoidal case (3 harmonics)}
	\label{tab:power_non_sinusoidal_3harm}
\end{table*}

The benefit of this proposal is evident when compensation scenarios are proposed. In this situation, certain specific targets are pursued, such as the minimisation of losses in power lines maintaining a constant active power flow $P$ or the elimination of harmonic currents generated by the load. In this situation, it is necessary to reduce the current to a minimum in such manner that it transfers the power to be converted into useful work. A compensator can be built with passive elements and active elements, such as controllable current sources that supply current or voltage sources that compensate distorted voltages. These elements can function independently or in a coordinated manner to form what is known as hybrid filters.

Now,  it is possible to identify which current components can be compensated, and through which elements, to maximise the power factor based on currents. Specifically, given the current defined in (\ref{eq:current_GAPoT}), the following can be shown:

\begin{itemize}
		\item A passive compensator composed of reactors can only compensate $\bm{I}_b$ by suitably choosing its susceptance $\bm{B}$.
		\item $\bm{I}_s$ and $\bm{I}_G$ can only be compensated by an active compensator based on nonlinear elements.
\end{itemize}

The formulation to carry out the compensation using passive elements is discussed in detail by the authors in \cite{montoya2019quadrature}.  Based on this work, new techniques can be developed for the definition of the reference current in active power filters and, consequently, to build more advanced compensation strategies. Most importantly, the techniques will be based on a complete and consistent power theory.          

\subsection{Circuits with harmonic generating (HG) loads}

\begin{figure}
	\centering
	\begin{circuitikz}[scale=1.5,/tikz/circuitikz/bipoles/length=1cm] 
		\draw
		(0,0) to[sV, v<=$u(t)$] (0,1)  to [ european resistor, l^=\small $R_s$] 
		(0,2)  to[short,i={$i$}] (3,2)
		(3,2) to [european resistor, l^=\small $R_l$]  (3,0) -- (0,0);
		\draw (-0.3,0.8) node{$+$};
		\draw (3,2) -- (4,2) to [american current source, l=$j_c$] (4,0) -- (3,0);
		\coordinate (left) at (-0.8,0);
		\coordinate (right) at (0.4,0);
		\coordinate (topl) at (-0.8,2);
		\coordinate (topr) at (0.4,2);
		\draw node[fit=(left)(right)(topl)(topr),draw, dashed, label={Source}, inner sep=10pt] {};		
		\coordinate (left2) at (2.9,0);
		\coordinate (right2) at (4.4,0);
		\coordinate (topl2) at (2.9,2);
		\coordinate (topr2) at (4.4,2);
		\draw node[fit=(left2)(right2)(topl2)(topr2),draw, dashed, label={Load}, inner sep=10pt] {};

		\draw (2.3,2) node[above left]{$x$};
		\draw (2.3,0) node[above left]{$x'$};
		\draw[<->,red] (2.3,0.1) -- (2.3,1.9) node[midway,fill=white,text=black] {$V_x$};
		
		\node[circle,draw=black, fill=white, inner sep=0pt,minimum size=5pt] (b) at (2.3,2) {};
			\node[circle,draw=black, fill=white, inner sep=0pt,minimum size=5pt] (b) at (2.3,0) {};
	\end{circuitikz}
	\caption{Non-linear load and real voltage source}
	\label{fig:non_linear_load}
\end{figure}
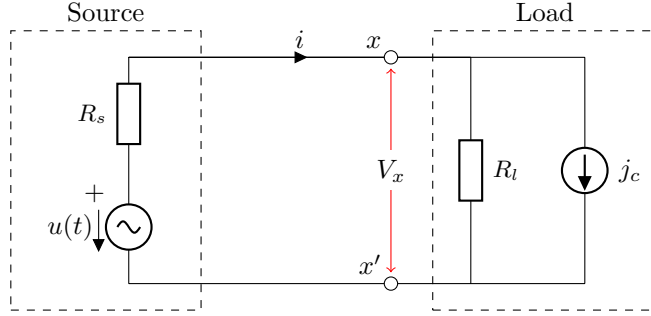

Finally, the ability of the proposed method to analyse electrical circuits with nonlinear loads is demonstrated. To do this, the circuit (see Figure \ref{fig:non_linear_load}) proposed by Czarnecki \cite{czarnecki1990powers} and analysed with GA by C-N in \cite{castro2010use} is revisited; the data show that the use of the present definition of apparent power $S$ leads to discrepancies in the concept of apparent power and reactive power.

The voltage and current source are as follows:

\begin{equation}
\begin{aligned}
	v(t)&=100\sqrt{2}\sin\omega t \\
	j_c(t)&=50\sqrt{2}\sin2\omega t
	\end{aligned}
\end{equation}

which are transformed to the geometric domain as:

\begin{equation}
\begin{aligned}
	\bm{V}&=-100\bm{\sigma}_{2} \\ 
	\bm{J}_c&=50\bm{\sigma}_{13}
	\end{aligned}
\end{equation}

Applying Kirchhoff's laws, setting $R_s=R_l=1\Omega$ and noting that there are no reactors, the following values are obtained:

\begin{equation}
\begin{aligned}
	\bm{V}_x&= -80\bm{\sigma}_{2}-40\bm{\sigma}_{13} \\
	\bm{V}_{R_s}&= -20\bm{\sigma}_{2}+40\bm{\sigma}_{13}\\
	\bm{I}&= -20\bm{\sigma}_{2}+40\bm{\sigma}_{13}\\
	\bm{I}_{R_l}&=-20\bm{\sigma}_{2}-10\bm{\sigma}_{13}
	\end{aligned}
\end{equation}

Applying the decomposition of currents suggested in this work, the total current $\bm{I}$ is as follows:

\begin{equation}
	\bm{I}=\bm{I}_a+\bm{I}_s+\bm{I}_b+\bm{I}_G=\underbrace{-20\bm{\sigma}_{2}}_{\bm{I}_a}+\underbrace{40\bm{\sigma}_{13}}_{\bm{I}_G}
\end{equation}

In Table \ref{tab:power_nonlinear}, the apparent geometric power of each element and the one that flows to the load is shown. The data show that the sum of the geometric power generated by the voltage source and that generated by the current source correspond to the geometric power consumed by the passive elements. In this case, each source generates $2000$ W of active power and $4000$ VA of nonactive power, but of opposite sign, thereby canceling the effects. The power that flows from the source to the load is that same non-active power, as reflected correctly by the term $\bm{M}_x$. Note that there is no geometric reactive power term ($\bm{\sigma}_{12}$). In addition, this verifies the determination of the flow direction of the net power, which in this case, is from the load to the source. 

\begin{table}[]
	\centering
	\begin{tabular}{@{}lrrr@{}}
		\toprule
		&  \multicolumn{2}{c}{\textbf{k-vector}} \\ \cmidrule{2-3}
		& \multicolumn{1}{c}{$\bm{\sigma}_0$} & \multicolumn{1}{c}{$\bm{\sigma}_{123}$} & \multicolumn{1}{c}{$\|\cdot\|$}   \\ \midrule
		$\bm{M}_u$    & 2000          & -4000      &         4472.1     \\ 
		$\bm{M}_{R_s}$    & 2000       &     &         2000.0   \\
		$\bm{M}_{R_l}$    & 2000       &      &   2000.0      \\  
		$\bm{M}_{J_c}$    & 2000     & 4000     & 4472.1   \\
		$\bm{M}_{x}$      &         & -4000     &   4000    \\ \bottomrule
	\end{tabular}
	\caption{Power decomposition non linear load in Figure \ref{fig:non_linear_load}}
	\label{tab:power_nonlinear}
\end{table}

As described above, it is possible to verify that our proposal is satisfied regardless of the value and order of the harmonics chosen. If, e.g., the value of the current source is $j_c(t)=50\sqrt{2}\sin3\omega t$, then the new power values are those shown in Table \ref{tab:power_nonlinear2}. The data show that the power balances continue to satisfy the PCoE, as expected. In this case, the flows change slightly to accommodate the change in current source. 

\begin{table}[]
	\centering
	\begin{tabular}{@{}lrrr@{}}
		\toprule
		&  \multicolumn{2}{c}{\textbf{k-vector}} \\ \cmidrule{2-3}
		& \multicolumn{1}{c}{$\bm{\sigma}_0$} & \multicolumn{1}{c}{$\bm{\sigma}_{1234}$} & \multicolumn{1}{c}{$\|\cdot\|$}   \\ \midrule
		$\bm{M}_u$    & 2000          & -4000      &         4472.1     \\ 
		$\bm{M}_{R_s}$    & 2000       &  -1600   &         2561.2   \\
		$\bm{M}_{R_l}$    & 2000       &   1600   &   2561.2       \\  
		$\bm{M}_{J_c}$    & 2000     & 4000     & 4472.1   \\
		$\bm{M}_{x}$      &         & -2400     &   2400    \\ \bottomrule
	\end{tabular}
	\caption{Power decomposition for non linear load with $j_c(t)=50\sqrt{2}\sin3\omega t$}
	\label{tab:power_nonlinear2}
\end{table}

\section{Conclusions}

This paper presents a completely major upgrade and reformulation for one of the main power theories
based on GA for single-phase circuits with linear and nonlinear loads under sinusoidal and nonsinusoidal
conditions. This new tool refines, corrects and improves the results obtained in previous studies. The
definition of geometric apparent power as the product of voltage and reversed current ensures a correct
determination of the flow of active power $P$ and nonactive power. Additionally, the optimal decomposition
of the load current into meaningfull engineering terms, enables the development of compensation strategies
not easily performed previously. The use of GA makes it possible to analyse electrical circuits in a unified
manner and in compliance with traditional laws that govern circuit theory. This work reiterates the concerns
raised by other renowned authors about the use of the traditional concept of apparent power $S$, since it is a controversial term that lacks a clear physical sense and that is not generally applicable in nonsinusoidal or
nonlinear systems. The use of this theory opens up new possibilities for the future analysis of new generation electrical systems, where the proliferation of nonlinear systems in the smart grid make the application of traditional power theories unfeasible. The development of the theory to multiphase system is a future task that will be performed thanks to de inherent multidimensionality of GA.

\section*{Acknowledgment}

\noindent This research has been supported by the Ministry of Science, Innovation and Universities at the University of Almeria under the programme  "Proyectos de I+D de Generación de Conocimiento" of the national programme for the generation of scientific and technological knowledge and strengthening of the R+D+I system with grant number PGC2018-098813-B-C33.

%\section*{References}
\bibliographystyle{elsearticle-num} 
\bibliography{mybibfile}

 \section*{Appendix}
\renewcommand{\thesubsection}{\Alph{subsection}}
\subsection{General concepts}
Given an ortho-normal base $\bm{\sigma}=\{\sigma_1,\sigma_2,...,\sigma_n\}$ for a vector space in $\mathbb{R}^n$, it is possible to define a new space called geometric algebra $\mathcal{G}_n$ whith a bilinear form. In this new space we  have vector bases $\bm{\sigma}$ and other type of elements of high dimensionality. For example, in the case of $\mathbb{R}^3$ (Euclidean space),  a basis composed by 3 unitary vectors $\bm{\sigma}=\{\sigma_1,\sigma_2,\sigma_3\}$ is defined. These unitary vectors fulfill that $\sigma_k\sigma_k=\sigma_k^2=1$ and $\sigma_i\wedge\sigma_j=0$ for $i\ne j$. Taking advantage of this property and using the Grassmann wedge product $\sigma_i \wedge \sigma_j = -\sigma_j \wedge \sigma_i$, it can be verified that

\begin{equation}
\begin{split}
(\sigma_i \wedge \sigma_j)^2 &=(\sigma_i\sigma_j)(\sigma_i\sigma_j)=\sigma_i(\sigma_j\sigma_i)\sigma_j=\sigma_i(-\sigma_i\sigma_j)\sigma_j= \\
&=-(\sigma_i)^2(\sigma_j)^2=-(1)(1)=-1
\end{split}
\end{equation}

It should be noted that $\sigma_i\sigma_j=\sigma_{ij}$ squares to $-1$, so we can conclude that we are dealing with a new element, namely a bivector. In the same way, the wedge product  of 3 vectors is called trivector, and in general, the product of $k$ vectors is called $k$-vector. It is therefore concluded that the most general basis for $\mathcal{G}_3$ is

\begin{equation}
\{1,\sigma_1,\sigma_2,\sigma_3,\sigma_{12},\sigma_{13},\sigma_{23},\sigma_{123}\}
\end{equation} 

Generally speaking, the elements of a geometric algebra are called multivectors $\bm{A}$ and are expressed as a linear combination of their different bases.

\begin{equation}
\bm{A} = \langle\bm{A}\rangle_0+\langle\bm{A}\rangle_1+\langle\bm{A}\rangle_2+...+\langle\bm{A}\rangle_n=\sum_{k=0}^{n}\langle\bm{A}\rangle_k
\end{equation} 

where each $\langle\bm{A}\rangle_k$ is an element of grade $k$, representing scalars (grade 0), vectors (grade 1), bivectors (grade 2) or in general, $k$-vectors (grade $k$).

\subsection{Geometric operations}

The geometric product is one of the fundamental contributions to the geometric algebra of Grassmann and Clifford. It is defined as the sum of the inner or scalar product and the wedge or Grassmann product. For the case of two vectors $\bm{a}$ and $\bm{b}$, it reads

\begin{equation}
\bm{a}\bm{b} = \bm{a}\cdot\bm{b}+\bm{a}\wedge\bm{b}
\end{equation} 

For the special case that the vectors are unitary bases $\sigma_i$ and $\sigma_j$ with $i \neq j$, the result obtained is a bivector.

\begin{equation}
\bm{\sigma}_i\bm{\sigma}_j = \bm{\sigma}_i\cdot\bm{\sigma}_j+\bm{\sigma}_i\wedge\bm{\sigma}_j=\bm{\sigma}_i\wedge\bm{\sigma}_j=\bm{\sigma}_{ij}
\end{equation}

\noindent In addition, since the wedge product is anticonmutative, we also have

\begin{equation}
\bm{\sigma}_{ij}=\bm{\sigma}_i\bm{\sigma}_j =\bm{\sigma}_i\wedge\bm{\sigma}_j=-\bm{\sigma}_j\wedge\bm{\sigma}_i=-\bm{\sigma_{ji}}
\end{equation}

Unlike vectors, whose square is 1, bivectors square to $-1$.

\begin{equation}
\bm{\sigma}_{ij}\bm{\sigma}_{ij} =\bm{\sigma}_i\bm{\sigma}_j\bm{\sigma}_i\bm{\sigma}_j=-\bm{\sigma}_j\bm{\sigma}_i\bm{\sigma}_i\bm{\sigma}_j=-\bm{\sigma}_j\bm{\sigma}_j=-1
\end{equation}

Finally, the important \textit{reversion} operation is defined. It applies to multivectors in the following way
\begin{equation}
\bm{A}^{\dagger}=\sum_{k=0}^{n}\langle \bm{A}^\dagger\rangle_k=(-1)^{k(k-1)/2}\langle \bm{A}\rangle_k
\end{equation}

The norm of a  multivector  $\|\bm{A}\|$ is always a scalar and can be obtained  as

\begin{equation}
\|\bm{A}\|=\sqrt{\left\langle A^\dagger A \right\rangle_0}=\sqrt{\left\langle A A^\dagger \right\rangle_0}=\sum_{k=0}^n{ \langle  \langle A \rangle_k \langle  A^\dagger \rangle_k \rangle_0}
\end{equation}

\end{document}